\documentstyle[emulateapj,onecolfloat]{article}

\newcommand{\tskip}{}%\vspace{2pt}}
\newcommand{\fsky}{f_{\rm sky}}
\newcommand{\simlt}{\lesssim}
\newcommand{\simgt}{\gtrsim}

\input epsf
\begin{document}
\twocolumn[
\title{Weak Lensing: Prospects for Parameter Estimation}
\author{Wayne Hu and Max Tegmark$^a$
}
\affil{Institute for Advanced Study, Princeton, NJ 08540}

\begin{abstract}
Weak lensing of galaxies by large scale structure 
can potentially measure cosmological quantities as precisely 
as the cosmic microwave background (CMB)
though the relation between the observables and the fundamental 
parameters is more complex and degenerate, especially in the full space of
adiabatic cold dark matter models considered here.  We
introduce a Fisher matrix analysis of the information contained
in weak lensing surveys to address these issues and provide
a simple means of estimating how survey propreties and source redshift uncertainties 
affect parameter estimation.    We find that even if the characteristic redshift of the
sources must be determined from the data itself, surveys on degree scales and above
can significantly assist the CMB on parameters that affect the growth rate of
structure.
\end{abstract}

]

\section{Introduction}

Weak lensing of faint galaxies by large scale structure can in principle provide
precise constraints on the spectrum and evolution of mass fluctuations in the universe 
(\cite{Mir91} 1991; \cite{Blaetal91} 1991; \cite{Kai92} 1992).  Given the same
sky coverage, the statistical errors on these measurements should be as 
small as those from 
the cosmic microwave background (CMB).
The main systematic errors are detector-based rather than astrophysical; though they
currently present a great obstacle against detection, it is one that is in principle
surmountable.   

Because lensing convolves aspects of the spectrum of present-day mass fluctuations,
their evolution, and the distribution of source galaxies, it is less
clear how to translate precision in the observables into precision in the
cosmological parameters.  Previous work has focussed on a relatively
small number of parameters such as the matter density and the present-day amplitude
of the its power spectrum assuming a fixed functional form 
and a fixed distribution of sources (e.g. \cite{JaiSel97} 1997; 
\cite{BerWaeMel97} 1997; \cite{Kai98} 1998). Even
so predictions depend strongly on prior assumptions for these parameters.  

In this {\it Letter}, we introduce a Fisher information matrix approach to assess the ability of weak-lensing power spectrum measurements to determine cosmological parameters. 
The virtue of this approach is its ability to simply quantify 
how assumptions about the survey properties, parameter space, fiducial model and 
prior knowledge from other cosmological measurements affect parameter estimation.
We study an 11 dimensional parameter space based on the adiabatic cold dark matter model
and show that information from CMB anisotropy measurements can be used in lieu of
large sky coverage in isolating several key cosmological parameters and measuring
the redshift distribution of the sources.  

We begin in \S \ref{sec:fisher} with the Fisher matrix formalism.  In \S \ref{sec:model},
we describe the parameterization of the cosmological model to which we apply this
formalism in \S \ref{sec:estimation}.  We study the effect of the source sampling
and distribution in \S \ref{sec:galaxy} and summarize our conclusions in 
\S \ref{sec:discussion}.

%\cite{SmaEllFit95,LupKai97,ForMelDan96}

\section{Fisher Matrix}
\label{sec:fisher}

By measuring the distortion of the shapes of galaxies due to the
tidal deflection of light by large scale structure, one can determine
the power spectrum of the convergence as a function of multipole 
or angular frequency on the sky $\ell$ (\cite{Kai92} 1992; 1998)
\begin{equation}
P_\psi(\ell) = \ell^4 \int d\chi {g^2(\chi) \over \sinh^6\chi}
P_\Phi(\ell/\sinh\chi, \chi) \,,
\label{eqn:ppsi}
\end{equation}
where $P_\Phi$ is the power spectrum of the Newtonian potential,
$\chi = H_0 \sqrt{1-\Omega_{\rm tot}}\int_0^z H^{-1} dz $
is the radial distance to redshift $z$ in curvature units, and
$g(\chi)$ weights the galaxy source distribution by the lensing
probability
\begin{equation}
g(\chi) = 2 \sinh \chi \int_\chi^\infty d\chi' n(\chi')
{\sinh(\chi'-\chi) \over \sinh(\chi')} \,,
\end{equation}
where $n(\chi)$ is the distribution of
sources normalized to $\int d\chi n(\chi) = 1$.  We use the \cite{PeaDod96} (1996)
scaling relation to obtain the non-linear density, and hence the potential power
spectrum.  Since we consider models with massive neutrinos,
we have replaced the linear growth rate found there with the scale-dependent growth
rates from \cite{HuEis98} (1998).

\begin{figure}[t]
\centerline{\epsfxsize=3.5truein\epsffile{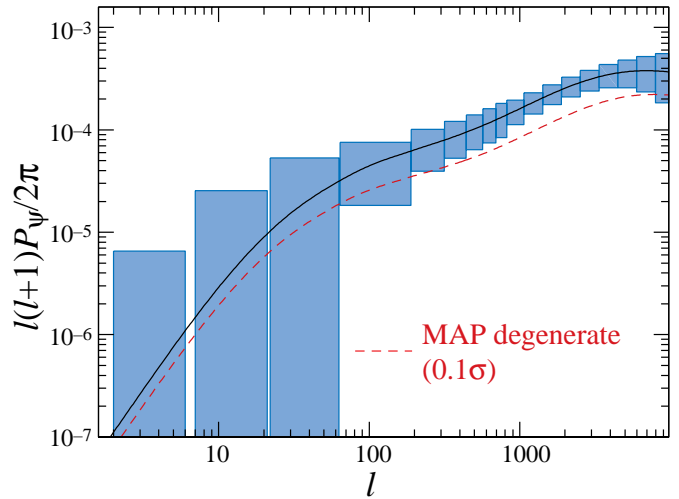}}
\caption{Weak lensing power spectrum for the fiducial $\Lambda$CDM model
and errors for a $\Theta_{\rm deg}=3$ survey.  These are compared
with a model that is  
degenerate with respect to CMB measurements from MAP.  
With the fixed lensing source distribution assumed here for
illustration purposes, the $0.1\sigma$ MAP separation increases to many $\sigma$ 
in the lensing survey.}
\label{fig:errorfig}
\end{figure}

\cite{Kai92} (1992,1998)
showed that the errors on a galaxy-ellipticity based estimator of 
$P_\psi(\ell)$ are described by
\begin{equation}
\Delta P_\psi = \sqrt{2 \over (2\ell +1)\fsky}(P_\psi + 4\left< \gamma_{\rm int}^2   
	\right>/\bar n) \,,
\label{eqn:deltaPpsi}
\end{equation}
where $\fsky = \Theta_{\rm deg}^2 \pi/129600 $ is the fraction of the
sky covered by a survey of dimension $\Theta_{\rm deg}$ in degrees and 
$\left<\gamma_{\rm int}^2\right>^{1/2} \approx 0.4$ is the intrinsic ellipticity of the
galaxies.  We assume throughout that $\bar n \approx 2 \times 10^{5} $deg$^{-2}
= 6.6 \times 10^{8}$sr$^{-1}$
which corresponds roughly to a magnitude limit of $R\sim 25$
(e.g. \cite{Smaetal95} 1995).   This sets the $\ell$ at which shot noise 
from the finite number of source galaxies becomes important.
The first term is simply the sampling error assuming gaussian statistics 
for the underlying field and makes the fractional errors of order unity
at the scale of the survey $\ell \sim 100/\Theta_{\rm deg}$.  
We plot an example of the band power $\ell(\ell+1)P_\psi/2\pi$ and its errors 
averaged over bands in $\ell$ in Fig.~\ref{fig:errorfig}.   

Equation (\ref{eqn:deltaPpsi}) tells us that in principle weak lensing can
provide measurements as precise as the CMB and has the benefit of being able
to probe significantly smaller angular scales.  
Unlike the CMB, the angular power spectrum of weak lensing
is rather featureless due to the radial projection in
equation~(\ref{eqn:ppsi}).
Thus the translation of
these measurements into cosmological parameters will suffer from more severe parameter
degeneracies.

To quantify this statement, we construct the Fisher matrix, 
\begin{equation}
{\bf F}_{ij} = -\left< \partial^2 \ln L \over \partial p_i \partial p_j \right>_{\bf x} \,,
\end{equation}
where $L$ is the likelihood of observing a data set ${\bf x}$
given the true parameters 
$p_1 \ldots p_n$. 
With equation~(\ref{eqn:deltaPpsi}), the Fisher
matrix for weak lensing becomes
\begin{equation}
{\bf F}_{ij} = \sum_{\ell=\ell_{\rm min}}^{\ell_{\rm max}}
	{\ell + 1/2 \over \fsky (P_\psi + 4\left< \gamma_{\rm int}^2 /\bar n\right>)^2} 
	{\partial P_\psi \over \partial p_i}
	{\partial P_\psi \over \partial p_j}\,.
\label{eqn:Fisher}
\end{equation}
Since the variance of an unbiased estimator of a 
parameter $p_i$ cannot be less than $({\bf F}^{-1})_{ii}$, the Fisher matrix
quantifies the best statistical errors on parameters possible with a given
data set.
 
We choose $\ell_{\rm min}=100/\Theta_{\rm deg}$ when evaluating 
equation~(\ref{eqn:Fisher}) as it corresponds roughly to the
survey size.  The precise value does not matter for parameter estimation due to the
increase in sample variance on the survey scale.  We choose a maximum value of
$\ell_{\rm max}=3000$ since here non-linear effects can produce non-gaussianity in
the angular distribution which increase the errors on the power spectrum
estimator (\cite{JaiSel97} 1997; \cite{JaiSelWhi98} 1998).  
Note that gaussianity is a better approximation for the shear field
than the density field due to the contribution of many independent lenses along
the line of sight. 
Again the exact 
value of the cutoff does not matter since shot noise begins to dominate 
at these scales (see Fig.~\ref{fig:errorfig}).   
Although information in the power spectrum is degraded by non-gaussianity, it
can be recovered from the non-gaussian measures such as the skewness
of the convergence.   We neglect such information here, but see \cite{JaiSel97} (1997) 
and \cite{BerWaeMel97} (1997).

\section{Parameterized Model}
\label{sec:model}

Projections for how well weak lensing can measure cosmological parameters
depend crucially on the extent of the parameter space considered
as well as the location in this space (or ``fiducial model'') around
which we quote our errors.   Previous works have focused on models
with essentially two parameters, the matter density $\Omega_m$ and
the amplitude of mass fluctuations on the 8 $h^{-1}$ Mpc scale today
$\sigma_8$ (e.g. \cite{BerWaeMel97} 1997, \cite{JaiSel97} 1997; \cite{JaiSelWhi98} 1998). 
Since all cosmological parameters that affect the amplitude of power
across a wide range of physical ($H_0 \simlt k \simlt 10 h $Mpc$^{-1}$)
and temporal scales ($z \simlt 1$) are accessible to weak lensing,
it seems prudent to consider a wider parameter space and then impose 
any external constraints as prior information.

We consider the adiabatic
cold dark matter model space and include 11 free parameters.
Weak lensing is only sensitive to 8 of the parameters:
the matter density $\Omega_m h^2$, the baryon density $\Omega_b h^2$,
the mass of the neutrinos $m_\nu$, the cosmological 
constant $\Omega_\Lambda$, 
the curvature $\Omega_K=1-\Omega_m-\Omega_\Lambda$, 
the scalar tilt $n_S$, 
the value of $P_\Phi(3000$ Mpc$^{-1}$) initially $A$, 
and the characteristic redshift of the sources $z_s$.  The other 3
parameters are necessary when considering prior
information provided by the CMB
and the galaxy power spectrum because of their covariance with
the 8 lensing parameters.  These are the optical
depth to reionization $\tau$, 
the primordial helium abundance $Y_p$, and the scalar-tensor
ratio $T/S$.

For our source redshift distribution we assume
a common redshift given by
$z_s$ since the errors on cosmological parameters are
insensitive to the shape of the distribution as long as it is 
considered known.  We return to this point in \S \ref{sec:discussion}.
Our fiducial model is the same $\Lambda$CDM model as chosen in \cite{EisHuTeg98}
(1998): $\Omega_m=0.35$, $h=0.65$, $\Omega_b=0.05$, $\Omega_\Lambda=0.65$, $m_\nu=0.7$ eV,
$\tau=0.05$, $n_S=1$, $T/S$=0, and $A$ given by the COBE normalization.  

\section{Cosmological Parameter Estimation}
\label{sec:estimation}

In Tab.~1, we present the Fisher estimates of errors on cosmological parameters
from weak lensing assuming full sky coverage $f_{\rm sky}=1$.  Errors for
more realistic sample sizes scale roughly as $f_{\rm sky}^{-1/2}$.  Although
these errors (per $f_{\rm sky}^{-1/2}$)
are comparable in precision to CMB estimates projected for
the MAP and Planck satellites from \cite{EisHuTeg98} (1998), 
they fail to achieve their ultimate potential due to parameter 
degeneracies.
We have included in parenthesis the degradation factors
due to degeneracies $(F_{ii})^{-1/2}/(F^{-1})_{ii}^{1/2}$.  These are on
the order of hundreds and represent the fact that lowering $\Omega_m h^2$ or
$n_S$ and
raising $\Omega_b h^2$ all reduce the primordial small-scale power in mass fluctuations
whereas raising $\Omega_\Lambda$, $\Omega_K$ or $m_\nu$ all slow the growth
rate of structure. These mimic changes in the amplitude $A$ and source redshift $z_s$
at the well-sampled high $\ell$'s.  
On the other hand the observables can basically be characterized by 4 parameters,
an amplitude, a slope, the non-linear scale ($\ell \sim 1000$) and the 
turnover scale ($\ell \sim 100$).

\begin{center}
{TABLE 1\\[4pt] \scshape 
Full-sky weak lensing survey compared with CMB satellites\footnote{Note that 
the MAP numbers assume temperature information 
only whereas the Planck numbers assume additional polarization information so as to
span the range of possible outcomes from the CMB missions. We also 
assume priors of $\sigma(Y_p)=0.02$ and $\sigma(z_s)=1$}
}\\[3pt]
\nopagebreak
\begin{tabular}{llll}
\tskip\tableline\tableline\tskip  $\sigma(p_i)$ &
 WL &  MAP & Planck  \\
\tskip\tableline\tskip\tskip
$\sigma(\Omega_m h^2)$ 
	&	0.024 (430)
	&	0.029 
	&       0.0027	\\
$\sigma(\Omega_b h^2)$
	&	0.0092 (310) 
	&	0.0029
	&       0.0002	\\
$\sigma(m_\nu)$
	&	0.29 (230)
	&	0.77
	&       0.25		\\
$\sigma(\Omega_\Lambda)$
	&	0.079 (180) 
	&	1.0
	&	0.11		\\
$\sigma(\Omega_K)$
	&	0.096 (200)
	&	0.29	
	&       0.030		\\
$\sigma(n_S)$
	&	0.066 (470) 
	&	0.1
	&       0.009 		\\
$\sigma(\ln A)$
	&       0.28 (310)	
	&	1.21
	&	0.045	\\
$\sigma(z_s)$	
	&	0.047 (56)		
	&	(1)
	&	(1)	\\

$\sigma(\tau)$
	&	--
	&	0.63
	&       0.004		\\
$\sigma(T/S)$
	&	--
	&	0.45	
	&	0.012 	\\
$\sigma(Y_p)$
	&	(0.02)    
	&	(0.02)		
	&	0.01		\\
\end{tabular}
\end{center}

Since surveys in the near future will be limited to several degrees on the
side at best ($f_{\rm sky} \sim 10^{-3}$), 
the precision lost to parameter degeneracies is crucial.  The
combinations of the parameters which are best constrained can
be determined by examining the eigenvectors of $F^{-1}$.   
The best constrained combination of parameters
involves $(\Omega_m h^2, \Omega_b h^2, \Omega_\nu h^2=m_\nu/94{\rm eV})$; 
variation in the direction $(-0.24, 0.45, 1)$ is constrained to have 
$1 \times 10^{-5}$ amplitude for $f_{\rm sky}=1$.  
Moving in this direction rapidly reduces the small scale power
in mass fluctuations and weak lensing is most sensitive to such
variations.  From analytic treatments
of growth rates, we also expect that neutrinos are twice as effective as baryons in
reducing small scale power (\cite{HuEis98} 1998).

These considerations imply that external constraints 
can help weak lensing measurements
regain their precision.
CMB satellite missions provide the ideal source of such information since 
the CMB angular power spectrum they measure is sensitive to the same 
cosmological parameters but in different combinations. 
In the example above, the CMB is particularly useful since it can provide
precise measurements of $\Omega_b h^2$ and $\Omega_m h^2$ leaving weak
lensing free to constrain the neutrino mass.
Furthermore, 
it is well known that CMB temperature measurements suffer from degeneracies
themselves, especially
between $\Omega_\Lambda$ and $\Omega_K$ along the direction that
keeps the angular diameter distance to last scattering fixed.
Because $\Omega_\Lambda$ must be raised to compensate $\Omega_K$ in
the CMB angular diameter distance, but must be lowered to compensate $\Omega_K$
in the growth rate of structure, one expects that weak lensing will
be particularly useful in breaking the degeneracy.

\begin{figure}[t]
\centerline{\epsfxsize=3.5truein\epsffile{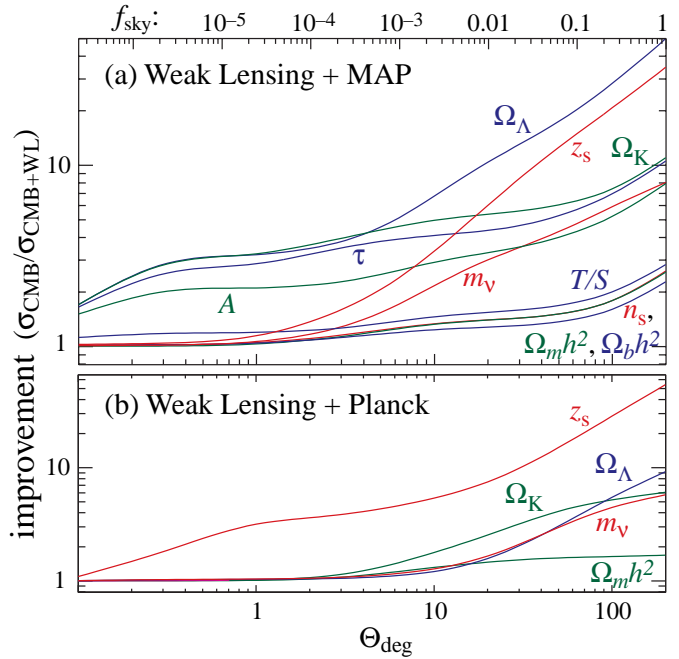}}
\caption{Improving CMB parameter estimation with weak lensing.  
We have assumed here and throughout a prior of $\sigma(z_s)=1$.}
\label{fig:improvement}
\end{figure}

Figure~\ref{fig:improvement} quantifies these expectations. The upper panel
shows the improvement over projected MAP satellite
errors on cosmological parameters (\cite{EisHuTeg98} 1998) 
when adding the weak lensing information with different survey sizes by summing
Fisher matrices.  By combining
these numbers with those of Tab.~1, one can read off the absolute errors
on cosmological parameters.  As expected, even a rather modest survey size of 
$\Theta_{\rm deg}=0.3$ is sufficient to improve MAP errors on $\Omega_\Lambda$
and $\Omega_K$ by a factor of 3 (see also Fig.~\ref{fig:errorfig}).   Ultimately,
weak lensing can improve MAP's measurement of these quantities by over an 
order of magnitude.
Amusingly, it also improves the measurement of $\tau$ by a comparable factor
since the angular diameter distance degeneracy in the CMB requires
$\tau$-variations to offset the 
amplitude changes from $\Omega_\Lambda$ and $\Omega_K$. 
Once the degeneracy is broken by weak lensing, $\tau$ becomes better measured.
With survey sizes of several degrees and beyond, constraints on $m_\nu$ improve
to reach the ultimate limit of $\sigma(m_\nu)=0.1$eV.

Weak lensing can improve on cosmological parameter estimation even if the CMB
reaches its full potential with precision temperature and polarization measurements
from the Planck satellite (see Fig.~\ref{fig:improvement}b).  In this case, gains will mainly come from survey
sizes $\Theta_{\rm deg}\simgt 10$.  Again there is the potential to improve
measurements of $\Omega_K$, $\Omega_\Lambda$ and $m_\nu$ by nearly an order
of magnitude, e.g. $\sigma(m_\nu)=0.04$eV.  This number is of particular interest
since the atmospheric neutrino anomaly is currently suggesting mass squared
separations of $\Delta m_\nu^2\sim 10^{-3}$.  More generally, this result
suggests that weak lensing and CMB measurements can be combined to study the
clustering properties of the dark matter and construct consistency tests 
that can confirm or rule out the presence of a cosmological constant as the
component which drives acceleration in the expansion rate.

\section{Galaxy Sampling and Distribution}
\label{sec:galaxy}

How does the sampling of galaxies and their redshift distribution affect 
parameter estimation?  \cite{Kai98} (1998) 
noted that at degree scales the large ratio of
sample variance to noise variance in equation~(\ref{eqn:deltaPpsi}) implies that one
can obtain better constraints on $P_{\psi}$ here by sparse sampling, provided the power
at smaller scales is measured to correct for the aliasing of power.  One can estimate
the gain in cosmological parameter estimation, under the optimistic assumption that
aliasing is negligible, by lowering the number density of galaxies $\bar n$
in equation~(\ref{eqn:deltaPpsi}) and 
computing the Fisher matrix as usual.  Unfortunately, going from a filled
$\Theta_{\rm deg}=3$ survey at $\bar n=2 \times 10^5$ deg$^{-2}$ to $\Theta_{\rm deg}=10$
at $\bar n=2 \times 10^{-4}$ deg$^{-2}$ not only does not improve the
errors, it can actually degrade them.   This is because the main source 
of cosmological information if $\Theta_{\rm deg}<10$ 
comes from the translinear regime near
$\ell=1000$.  Correspondingly in Fig.~\ref{fig:improvement}
parameter errors start improving rapidly only if $\Theta_{\rm deg}>10$
due to the resolution of the power spectrum bend below $\ell=100$.  Aliasing problems
may unfortunately preclude such agressive sparse sampling.

External knowledge of the redshift distribution of the sources can aid
parameter estimation especially for survey sizes with $\Theta_{\rm deg}<10$ where $z_s$
is not well-measured internally. Redshifts on a fair sample of 100 galaxies would be
sufficient to pin down the characteristic redshift to $\sigma(z_s)=0.1$, improving 
errors on $\Omega_\Lambda$ and $\Omega_K$
by up to a factor of two for $\Theta_{\rm deg}<10$.    Unfortunately spectroscopy on
a fair sample of these faint galaxies may be prohibitive.  Alternately
one can use photometric redshifts 
to select a subsample of galaxies whose individual redshifts are
known to $\sim 10\%$. 
For example, even separating the $1.3\%$ of galaxies that are at redshift 3
(\cite{Steetal96}\ 1996) 
improve errors on $\Omega_\Lambda$, $\Omega_K$, and $m_\nu$ 
by a factor of 2.

The actual value of the characteristic redshift itself also affects the sensitivity of weak lensing
to cosmological parameters but in a counterintuitive manner.   As the characteristic redshift of
the source galaxies rise, the lensing effect increases due to the increased amount 
of intervening large scale structure.   Though this makes the signal easier to detect,
it does not necessarily imply that
errors on cosmological parameters will improve.  In fact errors on $\Omega_\Lambda$
and $\Omega_K$
deprove as the redshift of the sources increases!  The reason is that
$\Omega_\Lambda$ and $\Omega_K$ only affect low redshift structure.  The intervening
high redshift structure is insensitive to these parameters but provide a larger
signal whose sample variance swamps the effect of $\Omega_\Lambda$ and $\Omega_K$.
By $z_s=3$, errors on $\Omega_\Lambda$ and $\Omega_K$ degrade by a factor of 6 and
2 for $f_{\rm sky}=1$.  

Finally, we have assumed that the galaxy redshift distribution is parameterized
by a single number, the characteristic redshift.
While this is indeed the main effect (\cite{Smaetal95b} 1995; \cite{ForMelDan96}
1995; \cite{LupKai97} 1997), 
the fact that weak lensing has the statistical power to measure the characteristic redshift
to better than $10\%$ for survey sizes $\Theta_{\rm deg}\simgt 10$, implies
that more detailed aspects of the distribution, e.g. its 
skewness, can in principle be measured from large surveys.   Allowing the data itself
to determine the form of the distribution will of course 
introduce more uncertainty in the cosmological pararameter determinations, but this
would be a small price to pay given the statistical power of such large surveys. 

\section{Discussion}
\label{sec:discussion}

The Fisher matrix analysis introduced here allows one to explore with ease 
how assumptions about the survey properties, the fiducial model and any 
prior knowledge from other cosmological measurements affect parameter estimation.
Weak lensing surveys are in principle sensitive to all cosmological parameters that
affect the shape of the matter power spectrum, the growth rate of fluctuations,
and the source redshift distribution.
Here we have included the effects of a cosmological constant, 
spatial curvature,
cold dark matter, baryonic dark matter, hot (neutrino) dark matter, power spectrum tilt
and amplitude, and the characteristic redshift of sources.
We find that even a relatively modest sample size of $0.3$ degrees would suffice
to improve our knowledge of cosmological parameters such as the cosmological
constant and the curvature over those provided by 
MAP satellite measurements of the CMB temperature power spectrum.  Order of magnitude
improvements in many cosmological parameters are available with survey sizes
$\simgt 3$ degrees.   

We have also explored how properties of the sample affect parameter estimation.
Sparse sampling can help extend power spectrum determinations to
larger angles but do not necessarily help parameter estimation due to the rather
featureless nature of the lensing power spectrum between $100 <\ell < 1000$.  The cosmological
constant and curvature can be best measured with a moderate redshift ($z_s \sim 1$) 
population of sources
since the larger signal at high redshifts is insensitive to these parameters and
act like noise for the purpose of determining those parameters. On the other hand,
separating out the $\sim 1\%$ of galaxies at $z_s \sim 3$ by photometric redshifts
and including this information in the analysis can assist determination of these 
parameters by up to a factor of 2.

The potential of weak lensing for cosmology explored here will only be realized
once systematic errors are reduced below the statistical errors
considered here.  An\-iso\-tropies in the point spread function of telescopes can 
mask the percent-level cosmological signal and pose a daunting challenge for the 
current generation of weak lensing surveys.   Our analysis reinforces the
conclusion that the returns for cosmology justify this great expenditure of effort. 

{\it Acknowledgements:} We thank R. Barkana, R. Blandford, D.J. Eisenstein, 
D. Hogg, M. Zaldarriaga for useful conversations.
W.H.\ is supported by the Keck Foundation, a Sloan Fellowship,
and NSF-9513835; M.T.\ by NASA through grant NAG5-6034 and Hubble
Fellowship HF-01084.01-96A from STScI operated by AURA, Inc.
under NASA contract NAS4-26555.

\end{document}